\documentclass[aps,pra,showpacs,twoside,twocolumn,longbibliography,10pt]{revtex4-2}
\usepackage[colorlinks=true, citecolor=blue, urlcolor=blue, linkcolor = blue ]{hyperref}
\usepackage{epsfig,newlfont,amssymb,amsfonts,amsmath,bm,subfigure,palatino,mathtools,amsthm,braket,times,soul,enumitem,color}
\usepackage[normalem]{ulem}
\newcommand{\stkout}[1]{\ifmmode\text{\sout{\ensuremath{#1}}}\else\sout{#1}\fi}
\usepackage[english]{babel}
\usepackage[utf8]{inputenc}
\usepackage{array}
\usepackage{xcolor}
\usepackage{graphics}
\newtheorem{theorem}{Theorem}

\def\Tr{\text{Tr}}

\usepackage{amsthm}
\usepackage{verbatim}
\usepackage{bbm}
\usepackage{wrapfig}

\usepackage{hyphenat}

\usepackage{float}

\newlength\figureheight 
\newlength\figurewidth 

\begin{document}

\title{Role reversal in quantum Mpemba effect}




\author{Arunabha Das}
\email{arunabha731234@gmail.com}

\author{Paranjoy Chaki} 
\email{pcapdj16@gmail.com}

\author{Priya Ghosh} 
\email{priyaghosh1155@gmail.com}

\author{Ujjwal Sen}
\email{ujjwal@hri.res.in}

\affiliation{Harish-Chandra Research Institute,  A CI of Homi Bhabha National Institute, Chhatnag Road, Jhunsi, Prayagraj  211 019, India}

\begin{abstract}

We investigate the quantum Mpemba effect in a dissipative Dicke model, 
which consists of a spin-1/2 ensemble coupled to a 
bosonic 
mode, which in turn is coupled to a bosonic bath.
We derive a sufficient criterion for  occurrence of the quantum Mpemba effect, characterized by quantum coherence,
in this model. 
We introduce the phenomenon of ``role reversal'' in the  Mpemba effect, wherein changes in the system parameters invert the relaxation ordering of a given pair of initial states that 
exhibit the Mpemba effect, causing the faster-relaxing state to become slower and vice versa.
We find the existence of role reversal in  Mpemba effect for this Dicke model using different relaxation measures, including differential quantum coherence and entanglement, and trace distance, between the time-evolved and steady states.
\end{abstract}

\maketitle
\section{Introduction}
The Mpemba effect (also known as the classical Mpemba effect)~\cite{Mpemba_1969,Jeng06,Katz_2009,Ibekwe_2016, classical_Mpemba_PNAS, Warring2024} describes an anomalous relaxation phenomenon in non-equilibrium classical systems, where a system initially farther from equilibrium relaxes faster than one closer to equilibrium. 
A well-known example is the cooling of water, where hot water reaches equilibrium sooner than the cold water.

In recent years, Mpemba-like behavior has been observed in quantum systems, giving rise to the quantum Mpemba effect (QME)~\cite{Ares_2025,mpemba-review-2}, which occurs in both closed and open quantum systems. In closed systems, the effect typically appears during symmetry restoration: two initially asymmetric states evolve under the same globally symmetric dynamics, and the state with greater initial asymmetry relaxes faster toward the symmetric configuration, as quantified by an asymmetry measure of the subsystem’s reduced density matrix~\cite{EA_1,EA_2,EA_3,EA_4_&_FD,EA_5,EA_6,EA_7,EA_8,EA_9,EA_10,EA_11,EA_12,EA_13,EA_14,EA_15,EA_16,EA_17,EA_18,EA_19,EA_20_&_Coherence_&_Ent_2,EA_21_&_Ent_3,Ent_4_&_density_matrix_element}. Whereas, in open quantum systems, QME refers to faster equilibration of states initially farther from equilibrium than those prepared closer to it.

In each (classical or quantum) Mpemba effect, an appropriate quantifier characterizing relaxation in nonequilibrium dynamics is required to identify the effect; for example, in the Mpemba effect observed in water cooling, temperature serves this role.
For open quantum systems, QME is most commonly characterized through the lens of distance-based measures
~\cite{Mpemba_Dicke_1_&_HS_dist_1,Tr_dist_2,KL_div_1,ED_1,Tr_dist_4,Tr_dist_5_&_KL_div_3_&_Entropic_distance_1,KL_div_2,KL_div_4,Tr_dist_6,L1_norm_dist_1,Tr_dist_7_&_HS_dist_3,Stat_dist_1,KL_div_5_&_total_variation_distance_1,HS_dist_2,norm_HS_dist_1,Tr_dist_3,Mpemba_Dicke_2_&_Tr_dist_1}.
Beyond distance-based measures, several quantum resource-based measures have also been employed to characterize QME in open systems, including the inverse participation ratio (IPR)~\cite{IPR_1}, quantum Fisher information \cite{QFI_1,QFI_2}, ergotropy~\cite{Ergotropy_1_&_Energy_1,Ergotropy_2}, and so on
~\cite{Ent_4_&_density_matrix_element,Energy_2,Temperature_1,Ent_1,Ergotropy_1_&_Energy_1,Coherence_Relative_Entropy,Order_parameter,EA_20_&_Coherence_&_Ent_2,EE_1,Ent_2,Krylov_Complexity,EA_21_&_Ent_3}. 

In this work, we study QME in the dissipative Dicke model, which consists of a spin-1/2 ensemble coupled to a bosonic mode, which in turn is coupled to a bosonic bath of harmonic oscillators.
To date, no criterion for observing QME in open systems formulated using quantum resource measures has been established. Here, we fill this gap by deriving a sufficient condition for observing QME, characterized by a quantum resource-based measure, differential quantum coherence, in the open Dicke model.

The occurrence of QME is highly sensitive to the choice of initial states, the system Hamiltonian, and, in open systems, the decay rates. 
The sensitiveness of occurrence of QME on the chosen initial states and the Hamiltonian parameters motivates us to introduce a phenomenon in QME, termed ``role reversal” in the quantum Mpemba effect. To describe the phenomenon, 
let us consider a relaxation quantifier such that two initial states are prepared with the same value of this measure at the initial time. If, for a given choice of Hamiltonian parameters, these states exhibit QME with respect to this measure, an intriguing phenomenon can arise. For the same pair of initial states, a change in the Hamiltonian parameters may cause their relaxation ordering to reverse: the state that previously relaxed faster becomes slower, and vice versa. This phenomenon is is referred to as the role reversal in QME, which highlights the strong dependence of the occurence of QME on the interplay between the chosen initial states and the Hamiltonian parameters.

Furthermore, we analyze whether the QME and its role reversal can be observed in the open Dicke model. We find the existence of both phenomena in this model using different classes of relaxation quantifiers, including quantum-resource-based measures such as differential quantum coherence and entanglement, as well as distance-based measures such as the trace distance between the time-evolved and steady states.

The rest of the paper is organized as follows.
In Sec.~\ref{sec-pre}, we briefly discuss QME, Markovian evolution of quantum states,  the dissipative Dicke model, and finally quantum resources, e.g., quantum coherence and entanglement.
In Sec.~\ref{sIII}, we provide a sufficient criterion for the occurrence of QME in the open Dicke model when differential quantum coherence serves as the relaxation measure.
In Sec.~\ref{sec-RR}, we introduce a phenomenon termed role reversal in QME, where a change in the Hamiltonian parameters alters the relaxation pathways of two evolving states, causing the state that initially relaxes faster in the QME to become slower, and vice versa.
Next, the occurrence of role reversal in QMEs for the open Dicke model characterized by differential quantum coherence, differential entanglement, and trace distance between the time-evolved and steady states are presented subsection-wise in Sec.~\ref{sec-RR}. Finally, we conclude in Sec.\ref{sec-Conclusion}.

\section{Gathering the tools}
\label{sec-pre}
In this section, we first present a brief overview of the QME. We then describe the Markovian dynamics and the associated vectorization method used to study the system’s evolution. Next, we describe the dissipative Dicke model and outline of the adiabatic elimination technique employed. Finally, we discuss the quantum resources that capture the signatures of the QME for the dissipative Dicke model in our analysis.

\subsection{Quantum Mpemba effect}
The classical Mpemba effect~\cite{Mpemba_1969,Jeng06,Katz_2009,Ibekwe_2016,classical_Mpemba_PNAS,Warring2024} is a counterintuitive phenomenon in which a system initially farther from equilibrium relaxes faster than one closer to equilibrium. This phenomenon was first observed~\cite{Mpemba_1969} in the cooling of water, where hot water freezes faster than cold water. 
Paradoxically, heating the water first can accelerate its subsequent freezing.
In line with the classical Mpemba effect, an analogous phenomenon has also been shown to exist in the quantum regime, known as quantum Mpemba effect~\cite{Mpemba_Dicke_1_&_HS_dist_1,Ares_2025,mpemba-review-2}. 
Let us consider that two systems are initially prepared in the states $\rho_0$ and $\rho'_0$, respectively, both evolving under the same dynamics. Their initial distances from the equilibrium state are $\mathcal{D}(\rho_0)$ and $\mathcal{D}(\rho'_0)$, with $\mathcal{D}(\rho'_0) < \mathcal{D}(\rho_0)$, where $\mathcal{D}(\cdot)$ denotes a distance measure between a given state ``$
\cdot$" and the steady state. If, at some finite time $t'$, the ordering reverses such that $\mathcal{D}(\rho'(t)) > \mathcal{D}(\rho(t))$ for $t > t'$, meaning that $\rho(t)$ reaches steady state faster than $\rho'(t)$ with respect to $\mathcal{D}(\cdot)$, then we conclude that the system exhibits the QME with respect to the distance measure $\mathcal{D}(\cdot)$. Here, $\rho(t)$ and $\rho'(t)$ denote the time-evolved states corresponding to the initial states $\rho_0$ and $\rho'_0$, respectively.

In scenarios where the two initial states are related by a unitary transformation, i.e., 
$\rho'_0 = U \rho_0 U^{\dagger}$, the Mpemba effect has been investigated using 
unitary-invariant distance measures such as the Hilbert-Schmidt distance~\cite{Mpemba_Dicke_1_&_HS_dist_1} and the trace distance~\cite{Mpemba_Dicke_2_&_Tr_dist_1}. In contrast to the conventional classical Mpemba 
effect, both dynamical trajectories originate from the same initial point, satisfying 
$\mathcal{D}(\rho_0) = \mathcal{D}(\rho'_0)$, yet one of the states relaxes to 
equilibrium faster than the other, spending a comparatively lesser amount of time.
In this work, we adopt the latter definition of the quantum Mpemba effect. However, rather than only restricting ourselves to the trace-distance measure from steady state as the sole indicator of QME, we additionally employ quantum differential coherence and entanglement as diagnostic quantities. Here, the differential coherence and entanglement denote the difference between the coherence and entanglement of the state at any given time and their respective steady values. Accordingly, for each of these measures, although different initial states begin with the same initial value, a more rapid relaxation toward the corresponding steady value constitutes clear evidence for the occurrence of the QME. Note that, in the case of entanglement and coherence, the initial states are related by local unitary transformations and coherence-preserving unitaries, respectively, thereby yielding identical initial values of entanglement and coherence.

\subsection{Evolution under Markovian dynamics}
Let us consider an initial state $\rho_0$ acting on the $d$-dimensional Hilbert space $\mathcal{H}_{d}$ and it evolves under a Markovian dynamics. The  evolution of the state can be
represented by Gorini-Kossakowski-Sudarshan-Lindblad (GKSL) master equation, i.e., $\dot{\rho}(t) = e^{t\mathcal{L}}[\rho_0]$ where the Liouvillian $\mathcal{L}$ is defined as \cite{Breuer_Petruccione_brief_notes,BreuerPetruccione2007,preskill2018quantum_ch3}
\begin{align}
    \mathcal{L}[X] = -\frac{i}{\hbar} [H,X] + \sum_{\mu=1}^{N_J}{(L_{\mu} X L_{\mu}^{\dagger} - \frac{1}{2} \{ L_{\mu}^{\dagger}L_{\mu} , X \})} \label{eq: Lindblad Master eqn}.
\end{align}
Here, $N_J$ is the number of Lindblad jump operators $L_{\mu}$.
The right eigenvectors $r_k$ and eigenvalues $\lambda_k$ of the Liouvillian $\mathcal{L}$ are defined as $\mathcal{L}[r_k] = \lambda_k r_k$. The left eigenvectors $\ell_k$ are defined as $\mathcal{L}^{\dagger}[\ell_k] = \lambda_{k}^{*} \ell_k$. The left and right eigenvectors hold the bi-orthogonality relation, $[\Tr(\ell_{j}^{\dagger}r_{k})]_{j \neq k} = 0$.
The state of the system at time $
t$ in a Markovian process is given by
\begin{equation}
    \rho (t) = \sum_{i=1}^{d^{2}}{e^{\lambda_{i}t}\frac{Tr(\ell_{i}^{\dagger}\rho_{0})}{Tr(\ell_{i}^{\dagger}r_{i})}r_{i}} .\label{Lindblad_Markovian_time_evolution}
\end{equation}

To evaluate the eigenvalues and left and right eigenvectors, we use the vectorization technique \cite{amshallem2015approachesrepresentinglindbladdynamics}. The Liouvillian ($\mathcal{L}$), written in Eq.~\eqref{eq: Lindblad Master eqn}, can be expressed in the vectorized form as follows:
\begin{align}
    & \mathcal{L}_{v} |X\rangle\rangle = [-\frac{i}{\hbar} (I \otimes H - H^T \otimes I)\\
    & + \sum_{\mu=1}^{N_J}{((L_{\mu}^{*} \otimes L_{\mu}) - \frac{1}{2}((I \otimes L_{\mu}^{\dagger} L_{\mu}) + (L_{\mu}^{\dagger} L_{\mu})^{T} \otimes I))})] |X\rangle\rangle \label{vec_liouv}.
\end{align}
For any arbitrary matrix $A$, $|A\rangle\rangle$ denotes the vectorized form of that matrix.
The right and left eigenvectors of the Liouvillian are denoted as $|r_k\rangle\rangle$ and $ |\ell_k\rangle\rangle$ respectively. They obey the following relations: $ \mathcal{L}_{v} |r_k\rangle\rangle = \lambda_{k} |r_k\rangle\rangle$ and $  \mathcal{L}_{v}^{\dagger}  |\ell_k\rangle\rangle = \lambda^{*}_{k} |\ell_k\rangle\rangle$ and their corresponding bi-orthogonality condition turns out to be $\langle\langle \ell_{j} | r_{k} \rangle\rangle_{j \neq k} = 0$.
The time-evolution of the state $\rho_0$ under the Markovian dynamics, governed by Liouvillian $\mathcal{L}$,  is given by ~\cite{Lindblad_analytical_vec_1,Lindblad_analytical_vec_2}
\begin{equation}
    |\rho(t)\rangle\rangle = \sum_{i=1}^{d^{2}}{e^{\lambda_{i}t}\frac{\langle\langle \ell_{i} | \rho_{0}\rangle\rangle}{ \langle\langle \ell_{i} | r_{i}\rangle\rangle} |r_{i}\rangle\rangle}\nonumber.\label{Lindblad_Markovian_time_evolution_vectorised}
\end{equation}
Here $Re(\lambda_{i}) \leq 0; \hspace{1 mm} \forall i \in \{1,...,d^{2}\}$ \cite{zhang2025directalgebraicproofnonpositivity} manifests the dissipative nature of the Markovian dynamics. As there is no dynamical time evolution once the steady state is reached, so $\mathcal{L}[\rho_{ss}] = \frac{d}{dt}(\rho_{ss}) = 0$, which can be rewritten as $\mathcal{L}[\rho_{ss}] = 0 \rho_{ss}$. Thus we see that $\rho_{ss}$ is the right eigenvector corresponding to zero eigenvalue of $\mathcal{L}_{v}$. We also see that $\mathcal{L}^{\dagger}[I_d]=0$, so $l_1 = I_d$ where $I_d$ denotes the identity operator on the Hilbert space of dimension $d$. Therefore, if we label the index in descending order according to the real parts of $\lambda_i$'s, i.e, $Re(\lambda_1) > Re(\lambda_2)> ... > Re(\lambda_{d^2})$, then $\lambda_{1}=0$ corresponds the steady state. When we consider that $\mathcal{L}$ has non-degenerate eigenvalues $\lambda_i$, the steady state, $\rho_{ss}$, is unique.
Now Eq.~\eqref{Lindblad_Markovian_time_evolution} can be rewritten as
\begin{align}
    \rho (t) = \rho_{ss} + \sum_{i=2}^{d^{2}}{e^{\lambda_{i}t}\frac{c_i}{k_i}r_{i}}, \label{Lind_Markov_Evo}
\end{align}
where $c_i \coloneqq \Tr(\ell_{i}^{\dagger}\rho_{0})$, $k_i \coloneqq \Tr(\ell_{i}^{\dagger}r_{i})$, and the steady state $\rho_{ss}$ is given by
\begin{equation}
    \rho_{ss} = \frac{1}{\Tr(r_{1})} r_1,  \label{rho_ss}
\end{equation}
since $ \ell_{1}^{\dagger}=I_{d}$ and $\Tr(\rho_0) = 1$.
Note that $\rho_{ss}$ is independent of the initial state. Therefore, the two initial states, $\rho_0$ and $\rho'_0$, lead towards the same steady state, $\rho_{ss}$, asymptotically under Markovian evolution, 
because $\lambda_i < 0; \hspace{1 mm} \forall i \in \{2,\ldots,d^2\}$.

\subsection{Dissipative Dicke model}
The Dicke model (closed)~\cite{Dicke_model_2,Mpemba_Dicke_1_&_HS_dist_1} describes the collective interaction between a bosonic mode, typically a cavity photon mode, and an ensemble of
$N$ two-level systems (atoms).
The Hamiltonian of the model is given by
\begin{equation}
    H = \Omega S_{z}+\omega a^{\dagger} a + \frac{g}{\sqrt{N}}(a+a^{\dagger}).\label{original_hamiltonian}
\end{equation}
The system comprises a bosonic harmonic oscillator, described by the Hamiltonian
$\omega a^{\dagger}a$, where $\omega$ denotes the angular frequency of the 
bosonic field. This mode interacts collectively with a spin system characterized by the
operator $S_z$, representing an ensemble of $N$ two-level quantum systems. Each constituent
two-level system is described by the local spin operator
$s_z^{(k)}=\tfrac{1}{2}\sigma_z^{(k)}$, where $\sigma_z^{(k)}$ is the Pauli matrix along the
$z$ direction and $S_z=\sum_ks_z^{(k)}$. The interaction between the collective spin and the bosonic mode is mediated
by the coupling term $\tfrac{g}{\sqrt{N}}(a+a^{\dagger})$, where g is the interaction strength. We consider that the parameters of the Hamiltonian, $\Omega$, $\omega$ and $g$, to be real numbers.

In this work, we consider a open quantum Dicke model~\cite{Dicke_model_1}, in which the system, described in Eq.~\eqref{original_hamiltonian}, is connected to a bosonic bath, consisting of quantum harmonic oscillators with $H_{bath}=\sum_{k}\omega_{k}b_{k}^{\dagger}b_{k}$ ~\cite{preskill2018quantum_ch3}. The interaction Hamiltonian between the system and bath is $\sum_{k}g_{k}(ab_{k}^{\dagger}+a^{\dagger}b_{k})$~\cite{preskill2018quantum_ch3}. 
The evolution of the system can be described in terms of the GKSL master equation with only one Lindblad jump operator, $L_{1}=\sqrt{k}a$ where $k$ is the decay rate and a positive real number~\cite{preskill2018quantum_ch3}.

Here, we focus on the largest symmetry sector~\cite{Mpemba_Dicke_1_&_HS_dist_1} where all spins are symmetrically aligned and it is the most collective behavior of the system. The highest possible total spin is $\frac{N}{2}$.
Therefore, $S_{z}|s,m_{s}\rangle = m_{s} \hbar |s,m_{s}\rangle$ and $m_{s}=-\frac{N}{2},-(\frac{N}{2}-1),...,+(\frac{N}{2}-1),+\frac{N}{2}$. Here, the dimension of the Hilbert space is, $d=2s+1=2\frac{N}{2}+1=N+1$. Now, $S^{2}|s,m_{s}\rangle=s(s+1)|s,m_{s}\rangle=\frac{N}{2}(\frac{N}{2}+1)|s,m_{s}\rangle$ which yields, $S^{2}=S_{x}^{2}+S_{y}^{2}+S_{z}^{2}=\frac{N}{4}(N+2)I$.
In this subspace, on the eigenbasis of $S_{z}$, $S_{z}=diag{(s,s-1,...,-s+1,-s)}$ and $\langle s, m_{i}|S_{x}|s,m_{j}\rangle = \frac{1}{2}(\sqrt{s(s+1)-m_{j}(m_{j}+1)}\delta_{m_{i},m_{j}+1} + \sqrt{s(s+1)-m_{j}(m_{j}-1)}\delta_{m_{i},m_{j}-1})$.

\subsubsection{Adiabatic elimination of the bosonic mode}
 An effective, purely spin-based representation of the Dicke model can be obtained by adiabatically eliminating the bosonic mode~\cite{Adiabatic_elimination_1,Adiabatic_elimination_2,Mpemba_Dicke_1_&_HS_dist_1}, which imposes the constraint that the decay timescale of the bosonic mode is much faster than the dynamics of the spin degrees of freedom.

 The Heisenberg equation of motion for the annihilation operator, in the open Dicke model, is given by
\begin{align*}
    \frac{da}{dt}=\mathcal{L}^{\dagger}[a]
    =(-i\omega-\frac{k}{2})a-i\frac{g}{\sqrt{N}}S_{x}.
\end{align*}
The adiabatic elimination approximation implies that, $\frac{da}{dt}=0 \Rightarrow a=-\frac{g(4\omega+2ik)}{\sqrt{N}(4\omega^{2}+k^{2})}S_{x}$. Substituting this expression of the creation operator and annihilation operator $a^{\dag}$ and $a$ respectively, into Eq.~\eqref{original_hamiltonian}, we obtain that the Dicke Hamiltonian reduces to~\cite{Mpemba_Dicke_1_&_HS_dist_1}
\begin{equation}
    \tilde{H}=\Omega S_{z}-\frac{4 \omega g^{2}}{\sqrt{4\omega^{2}+k^{2}}}\frac{S_{x}^{2}}{N} \label{H_eff},
\end{equation}
and the Lindblad jump operator $L_{1}=\sqrt{k}a$ becomes
\begin{equation}
    \tilde{L}_{1}=-\frac{g\sqrt{k}(4\omega+2ik)}{\sqrt{N}(4\omega^{2}+k^{2})}S_{x} \hspace{0.20cm}
\label{eff_jump_operator}.
\end{equation}

\subsection{Quantum resources}
{A quantum resource is a feature of a quantum system that provides an advantage over classical systems for performing certain tasks and cannot be created freely under a specified set of allowed operations, where the operations are known as free operations~\cite{Resource_Theory_1,Resource_Theory_2}.}
Quantum coherence and entanglement act as a resource in a variety of quantum information processing tasks and quantum technologies
~\cite{Ekert_protocol,Dense_coding,DJ_algo,teleportation,DC_tele_Ekert}.

\textbf{Quantum coherence:} 
Quantum coherence~\cite{coherence-review} is a fundamental property of quantum systems that arises from the principle of superposition. Unlike many other quantum resources, coherence is inherently basis dependent. A widely used and faithful quantifier of quantum coherence is the $l_1$-norm of coherence~\cite{PhysRevLett.113.140401}, which, for an arbitrary quantum state $\rho$ expressed in a given basis, is defined as
\begin{equation}
    l_{1}(\rho) = \sum_{j \neq k} |\rho_{jk}| ,
    \label{l1_norm}
\end{equation}
where $\rho_{jk}$ denotes the $(j,k)$-th off-diagonal element of the density matrix $\rho$ in that basis. Throughout our work, we measure the quantum coherence in the computational basis.

\textbf{Entanglement:}~
Entanglement~\cite{ent-review} is a quantum correlation that may be present in bipartite or multipartite systems. There are so many measures of bipartite entanglement, such as von Neumann entropy~\cite{Quantifying_entanglement}, relative entropy of entanglement~\cite{Quantifying_entanglement}, logarithmic negativity~\cite{log_neg_1,log_neg_2}, etc. In our work we consider logarithmic negativity as a faithful measure of bipartite entanglement.   
 Let us consider a bipartite system in a state $\rho_{AB}$ on the composite Hilbert space $\mathcal{H}_{A}^{d} \otimes \mathcal{H}_{B}^{d}$.
The negativity $\mathcal{N}(\rho_{AB})$~\cite{log_neg_1, log_neg_2} for a bipartite state $\rho_{AB}$, is the absolute value of the sum of all the negative eigenvalues of the partially transposed state, $\rho_{AB}^{T_A}$ of $\rho_{AB}$, where the partial transposition is
being taken with respect to the subsystem $A$, and is given by
\begin{align*}
    \mathcal{N} (\rho_{AB}) = \frac{|| \rho_{AB}^{T_A} || - 1}{2},
\end{align*}
where $||\rho||_{1}=\Tr{\sqrt{\rho^{\dagger} \rho}}$ is the trace-norm of the matrix $\rho$.

The logarithmic negativity ($\mathcal{L}_{N}$) \cite{log_neg_1,log_neg_2}, is defined in terms of negativity, is given by
\begin{align*}
    \mathcal{L}_{N}(\rho_{AB}) = \log_{2} [ 2 \mathcal{N}(\rho_{AB}) + 1].
\end{align*}
The nonzero value of $\mathcal{L}_{N}$ for any system, ensures the presence of non-vanishing bipartite entanglement in the system.

\section{Slowing of decoherence}
\label{sIII}
In this section, we demonstrate the existence of the QME in the Dicke model, characterized through differential quantum coherence, and derive a sufficient criterion for its occurrence.
In this regard, we consider a pair of initial states $\{\rho_0,\rho'_0\}$, where $\rho'_{0}=U(\beta)\rho_{0} U(\beta)^{\dagger}$. Here, $U(\beta) (\coloneqq e^{i \beta \sigma_{z}})$ with $\beta \in [0,2\pi)$, is a coherence-preserving unitary, which preserves the coherence in the eigenbasis of $\sigma_z$
~\cite{coherece_preserving_unitary_z_basis_qubit}. $\sigma_{z}$ is the Pauli spin matrix along the $z$ direction. The coherence of $\rho_0$ and $\rho'_0$ in the computational basis is same, i.e., $l_{1}(\rho_0) = l_{1}(\rho'_0)$.
We analytically derive a sufficient criterion under which $\rho'_0$ decoheres more slowly than $\rho_0$ over time under the Markovian dynamics of disspative Dicke model, where decoherence stands for the decrease of coherence. Therefore, the two states follow distinct relaxation pathways and require different durations to reach the zero differential coherence value asymptotically when the condition is satisfied by system parameters, thereby exhibiting QME by the set of states $\{\rho_0,\rho'_0\}$.

Note that, in the case of Dicke model, the steady state, $\rho_{ss}$, is diagonal on the eigenbasis of $S_z$, i.e., $(\rho_{ss})_{jk} = 0$ for $j \neq k$, shown in Appendix.~\ref{Ap: proof}. 
As in our case, coherence of the steady state itself is zero in the computational basis (discussed in~\ref{Ap: proof}), instead of the differential coherence, we only focus on the dynamics of the coherence of the state pair $\rho_0$ and $\rho'_0$ solely.
It can be shown that coherence of the evolved state $\rho(t)$ at any time can be written as (for detailed derivation, see Appendix.~\ref{Ap: l1_norm})
\begin{equation}
    l_{1}(\rho (t)) = \sum_{j \neq k}{|\sum_{i=2}^{d^{2}}{e^{\lambda_{i}t}\frac{c_{i}}{k_{i}} (r_{i})_{jk}}|}. \label{coh}
\end{equation}
Eq.~\eqref{coh} shows that $l_{1}(\rho(t))$ is a monotonically decreasing function of time under the Markovian evolution. $l_{1}(\rho(t))$ reaches towards steady state asymptotically which has a zero quantum coherence in the eigenbasis of $S_z$.
\begin{figure}
    \includegraphics[width=9cm]{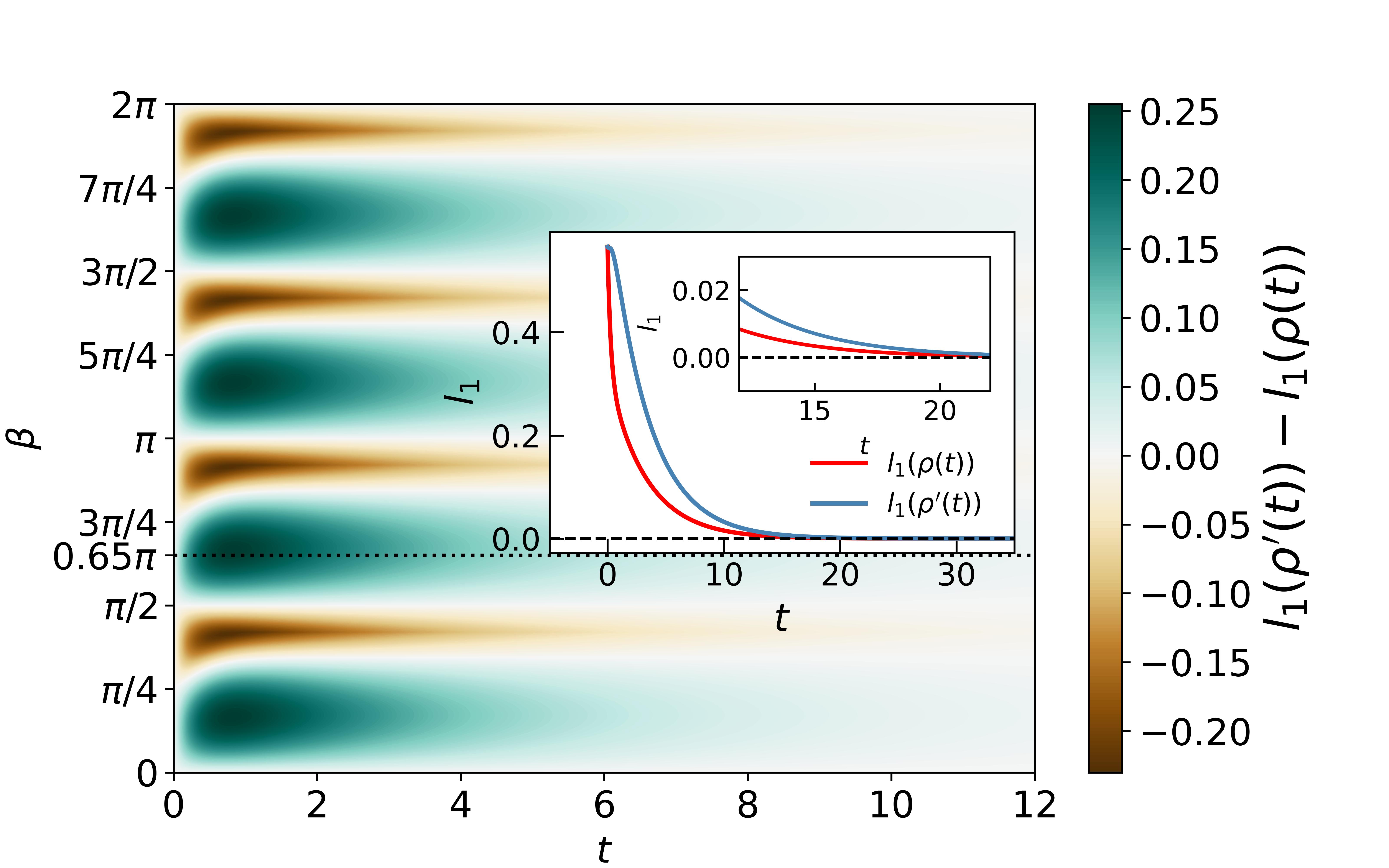}
\caption{\textbf{Quantum Mpemba effect characterized by differential coherence.}
We plot a heat map of $l_{1}(\rho'(t))-l_{1}(\rho(t))$  as a function of $t$ and $\beta$. We consider a qubit state represented in Bloch sphere representation, with $r_x = r_y = 0.4$ and the model parameters $\Omega= \omega = k =1$, $g=3$. We find that $l'_{1}(\rho(t))-l_{1}(\rho(t))>0$, for all finite $t \in (0,\infty)$ where $\beta \in (0, \frac{\pi}{4}] \cup (\frac{\pi}{2},\frac{3\pi}{4}] \cup (\pi,\frac{5\pi}{4}] \cup (\frac{3\pi}{4},\frac{7\pi}{4}]$, manifesting that the initial state obtained after the action of the unitary operator $U(\beta)$ decoheres more slowly towards the zero coherence state. Here, the time is described in the unit of $\frac{\Omega}{\hbar}$ whereas $\beta$ is dimensionless. At sufficiently large time, i.e., $t \rightarrow \infty$, $l'_{1}(\rho(t))-l_{1}(\rho(t)) = 0$.  In the inset, we plot $l_1$-norm measure of coherence with time $(t)$, for $\beta=0.65 \pi$, and we observe that the coherence of $\rho'(t)$ reaches zero slower in comparison to that of $\rho(t)$, exhibiting QME. 
}
    \label{fig:suff}
\end{figure}



Any general qubit state can be written as
\begin{align}
    \rho_{0} &= \frac{1}{2}(I_{2}+\vec{r}.\vec{\sigma})\label{state_Bloch_sphere},
\end{align}
where $\vec{r} = r_x\hat{x} + r_y\hat{y} + r_z\hat{z}$ is called the Bloch vector where $r_x,r_y,r_z \in [-1,+1]$ and $r_x^2 + r_y^2+r_z^2 \leq 1$.

We evaluate $c'_i \coloneqq \Tr(\ell_{i}^{\dagger} \rho'_0)$, in terms of the parameters of $\rho_{0}$ and $U(\beta)$, and obtain (for detailed derivation, see Appendix.~\ref{Ap: c_relation})
\begin{align}
    c'_{i}=c_{i} + \sin{\beta} [(\alpha_{x})_{i}(r_{y}\cos{\beta}-r_{x}\sin{\beta}) \nonumber\\
    - (\alpha_{y})_{i}(r_{y}\sin{\beta}+r_{x}\cos{\beta}))], \label{cpi}
\end{align}
where $(\alpha_{x})_{i} \coloneqq \Tr(\ell_{i}^{\dagger}\sigma_{x})$ and $(\alpha_{y})_{i} \coloneqq \Tr(\ell_{i}^{\dagger}\sigma_{y})$.
Let us consider $r_{x}=r_{y}$ as a special case that simplifies Eq.~\eqref{cpi} as follows:
\begin{equation}
    c'_{i}=c_{i} + r_{x} \sin{\beta} [(\alpha_{x})_{i}(\cos{\beta}-\sin{\beta}) - (\alpha_{y})_{i}(\sin{\beta}+\cos{\beta}))] . \label{cpi_s}
\end{equation}

Furthermore, note that Eq.~\eqref{cpi} depends only on $r_x$ and $r_y$ and is independent of $r_z$. Consequently, if the initial states lie on the $z$ axis of the Bloch sphere ($r_x=r_y=0$), one has $c_i’=c_i$, which implies $l_1(\rho(0))=l_1(\rho’(0))$. 
Since the coherence of the evolved state, $l_{1}(\rho (t))$, depends on the initial state through $c_i$ only and these coefficients remain equal for both states, it follows that $l_1(\rho(t))=l_1(\rho’(t))$ for all $t\geq 0$. Therefore, no QME occurs in this case.

Let us now state a sufficient criterion on the occurrence of the QME in the dissipative Dicke model.
\begin{theorem}
    \label{Theorem: suff_cond}
    Let $\{\rho_0,\rho_0'\}$ be two initial states evolving under the dissipative Dicke model, parameterized by $\{\Omega, \omega, g, k\}$ with $\Omega=\omega=k\coloneqq p$, where $p \in \mathbb{R}^+$ and $g \in \mathbb{R}$. Here, 
    $\rho_0$ is any general qubit state with Bloch vector satisfying $r_x=r_y$ 
    and $\rho'_0 \coloneqq U(\beta)\rho_0 U^{\dagger}(\beta)$, where $U(\beta) \coloneqq e^{i\beta\sigma_z}$ is a coherence-preserving unitary with $\beta \in [0,2\pi)$.
    The Mpemba effect, characterized by differential quantum coherence, is manifested in this Dicke model, independently of the values of $r_x=r_y$ and $r_z$,
    if the following two conditions hold: (1) $g>\sqrt{5}p$,  and (2) $\beta \in (0,\frac{\pi}{4}] \cup (\frac{\pi}{2},\frac{3\pi}{4}] \cup (\pi,\frac{5\pi}{4}] \cup (\frac{3\pi}{4},\frac{7\pi}{4}]$.
\end{theorem}

The proof of Theorem.~\ref{Theorem: suff_cond} is provided in Appendix.~\ref{Ap: proof}. 
In Fig.~\ref{fig:suff}, we plot $l_{1}(\rho'(t))-l_{1}(\rho(t))$ with respect to $t$ and $\beta$ for the numerical verification of Theorem.~\ref{Theorem: suff_cond}. Here, the time is described in the unit of $\frac{\Omega}{\hbar}$. Here, we consider a qubit state represented in Bloch sphere representation, with $r_x = r_y = 0.4$, and the parameters $p=1$, $g=3$. As the initial states are connected by a coherence-preserving unitary operation, we obtain $l_{1}(\rho'_0)-l_{1}(\rho_0)=0$, is represented by the white vertical straight line at $t=0$. Owing to decoherence during the evolution toward the zero-coherence state, the quantity
$l_{1}(\rho'(t)) - l_{1}(\rho(t))$ vanishes in the long-time limit, i.e., $t \rightarrow \infty$. This behavior is illustrated by
the fading of the color intensity toward white along the $t$-axis. The heat map in Fig.~\ref{fig:suff} spectacularly shows that $l_{1}(\rho'(t))-l_{1}(\rho(t)) > 0$, for all finite $t \in (0, \infty)$ and $\beta \in (0, \frac{\pi}{4}] \cup (\frac{\pi}{2},\frac{3\pi}{4}] \cup (\pi,\frac{5\pi}{4}] \cup (\frac{3\pi}{4},\frac{7\pi}{4}]$. Now, $l_{1}(\rho'(t))-l_{1}(\rho(t))>0$ yields $l_{1}(\rho'(t))>l_{1}(\rho(t))$ which is sufficient to confirm that the time-evolved state $\rho'(t)$ of the initial state, $\rho'_0$ which we obtain after the action of the unitary operator, decoheres more slowly towards the zero-coherence steady value. In the inset, we plot $l_1$-norm measure of coherence with time $t$, for $\beta=0.65 \pi$. The red and blue smooth curves represent the dynamics of $\rho(t)$ and $\rho'(t)$, the dynamics of the initial states $\rho_0$ and $\rho'_0$ respectively. Although the initial states $\rho_0$ and $\rho'_0$ have the same coherence due to the coherence-preserving unitary operator, the coherence of $\rho'(t)$ reaches zero slower, spending much more time in comparison to that of $\rho(t)$, up to some numerical precision. Thus, the condition on the parameters of the setup, the coherence-preserving unitary operator, and the initial states plays the role of a sufficient condition to yield the QME using differential quantum coherence for the open Dicke model.

\begin{figure}
    \centering
    \includegraphics[
        width=8.7cm
    ]{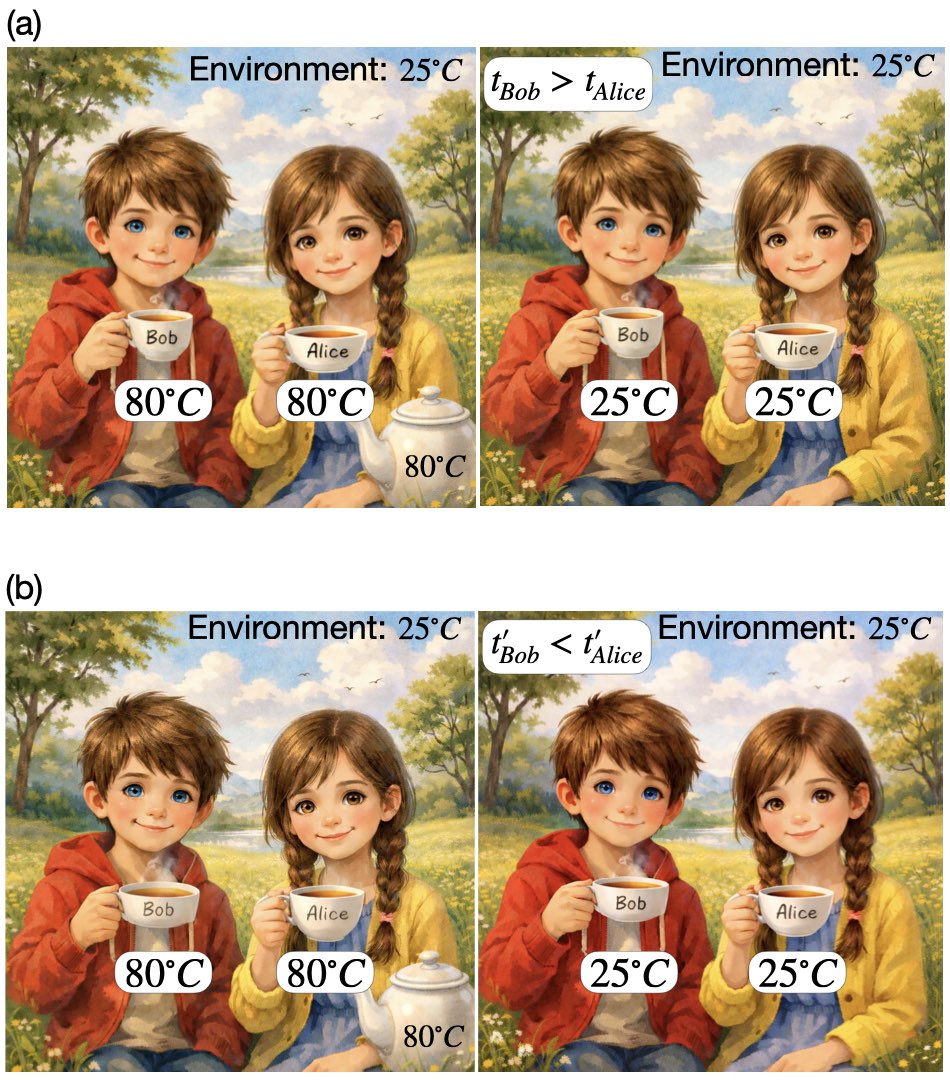}
    \caption{\textbf{Schematic representation of the quantum Mpemba effect and role reversal.} 
    Let us consider a situation when Alice and Bob have two cups of tea, each with the same amount. The initial (at $t=0$) temperature of the tea in both cups is same, i.e., $80^{\circ} C$ and both the cups are in the same environment with temperature $25^{\circ} C$. Alice and Bob choose their cups with two different shapes with different opening areas. The tea in the flat cup reaches the environment temperature $25^{\circ} C$ faster than the other. We define this situation as QME in our context. Let us consider two situations now. Panel~(a) depicts a situation when Alice chooses a flat cup while Bob does not. In this situation, if the tea Alice has spends $t_{Alice}$ amount of time while the tea Bob has spends $t_{Bob}$ amount of time before reaching the environment temperature $25^{\circ} C$, then $t_{Bob} > t_{Alice}$, exhibiting the QME according to the definition. 
    In contrast, Panel~(b) shows another situation when Bob chooses a flat cup while Alice does not, unlike the previous situation. In this situation, if the tea Alice has spends $t'_{Alice}$ amount of time while the tea Bob has spends $t'_{Bob}$ amount of time before reaching the environment temperature $25^{\circ} C$, then $t'_{Bob} < t'_{Alice}$. This also manifests the QME, but the roles of the two dynamics have been reversed. Here, the temperature, choices of cups, and the tea Alice and Bob possess, are analogous to the distance or quantity $\mathcal{M}$, the tuning parameters $\{\alpha_i\}$ and time-evolved pair of the initial states $\{\rho_0,\rho'_0\}$, respectively. Thus, Panel~(a) and Panel~(b) exhibit two situations of the QME with reversed roles w.r.t. each other.}
    \label{fig:Schematic}
\end{figure}

\section{Role reversal in  quantum Mpemba effect}
\label{sec-RR}
\begin{figure*}
    \centering
    \includegraphics[
        width=18cm,height=11cm,
    ]{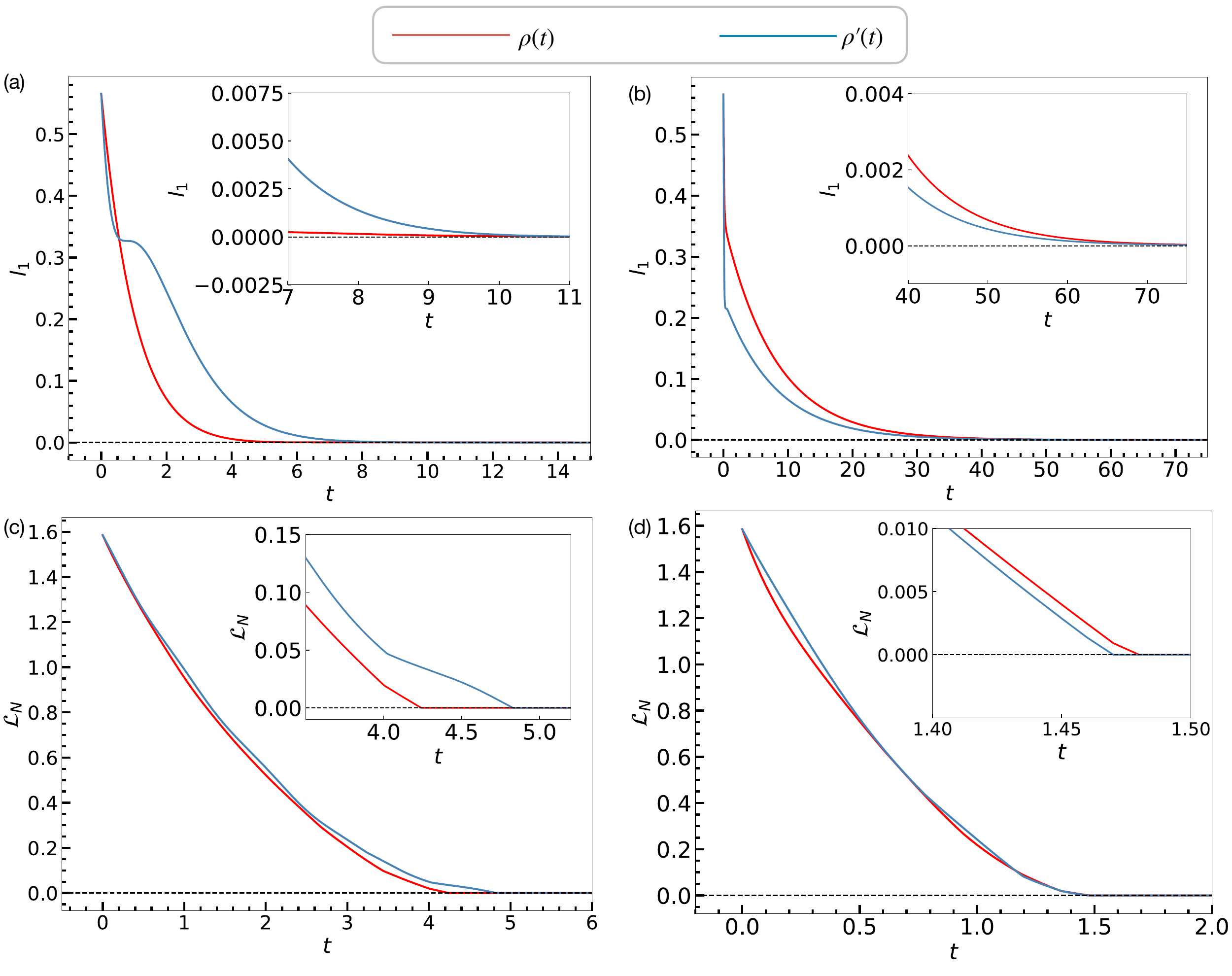}
    \caption{\textbf{Mpemba effect and its role reversal, characterized by differential quantum coherence and entanglement}.
    Panels (a) and (b) correspond to the dynamics of differential quantum coherence, keeping $N=1$, for the pair of initial states $\{\rho_0,\rho'_0\}$ for different parameter values of the Hamiltonian, i.e., $\{g=1$, $\omega = 0.1\}$ and $\{g=4.5, \omega = 1\}$ respectively. The other parameters of the Hamiltonian, the parameters of the initial state $\rho_0$ and the coherence-preserving unitary operation $U(\beta)$ are chosen as $r_{x}=r_{y}=0.4$, $\Omega = 1$, $k = 1$, $\beta = 0.33 \pi$. The smooth red and blue lines, in both sub-figures (a) and (b), correspond to the dynamics of coherence of $\rho(t)$ and $\rho'(t)$ corresponding to the initial states $\rho_0$ and $\rho'_0$, respectively. It is worth noting that both of the panels (a) and (b) manifest the QME characterized by differential quantum coherence, with reversed roles w.r.t. each other.
    Similarly, the panels (c) and (d) correspond to the dynamics of entanglement for the pair of evolved states $\{\rho(t),\rho'(t)\}$ corresponding to the initial states $\{\rho_0,\rho'_0\}$ respectively, for different parameter values of $\omega_{B}$, i.e., $8.88$ and $2.79$ respectively. The other parameters of the Hamiltonian are $N_{A}=N_{B}=3$, $\Omega_{A} = 3$, $g_{A} = 1$, $k_{A}=1$, $\omega_{A}=1$, $\Omega_{B} = 2.5$, $g_{B} = 3.5$ and $k_{B}=3$. The smooth red and blue lines, in both sub-figures (c) and (d), correspond to the dynamics of entanglement corresponds to $\rho_0$ and $\rho'_0$  respectively. Note that both of the panels (c) and (d) illustrate the QME characterized by differential entanglement, with reversed roles w.r.t. each other. In all the sub-figures, the time is described  in the unit $\frac{\Omega}{\hbar}$.}
    \label{fig:RR_plots_resources}
\end{figure*}

Here, we introduce a phenomenon, referred to as ``role reversal" in the quantum Mpemba effect.

Considering a pair of unitarily connected initial states $\{\rho_0, \rho'_0\}$ such that,
for any measure of distance from the steady state or any differential quantity $\mathcal{M}$
satisfying $\mathcal{M}(\rho'_0) = \mathcal{M}(\rho_0)$, there exists a specific set of
Hamiltonian's parameters $\{\alpha_1, \alpha_2, \ldots, \alpha_n\}$ for which the
evolution of $\mathcal{M}(\rho(t))$ approaches the asymptotic steady value faster than
that of $\mathcal{M}(\rho'(t))$, where $\rho(t)$ and $\rho'(t)$ are the time-evolved states corresponding to the initial states $\rho_0$ and $\rho'_0$ respectively. Therefore, the QME is captured by
the differential quantity $\mathcal{M}$.
However, if we keep the pair of initial states $\{\rho_0, \rho'_0\}$ fixed and change the parameters of the Hamiltonian, there exists some set of parameter values $\{\alpha'_1,\alpha'_2,...,\alpha'_n \}$ for which the dynamics of $\mathcal{M}(\rho(t))$ becomes slower than that of $\mathcal{M}(\rho'(t))$, in reaching towards the steady zero value. This scenario also exhibits the QME, albeit in a reversed form. In the former
case, $\mathcal{M}(\rho_0(t))$ relaxes to the steady value faster, whereas in the latter
case, $\mathcal{M}(\rho'_0(t))$ does so more rapidly. 
Consequently, the roles of the
dynamical evolutions of $\mathcal{M}$ corresponding to the two initial states $\rho_0$ and $\rho'_0$ are reversed. We refer to this phenomenon as a role reversal in quantum Mpemba effect.

In the following three subsections, we demonstrate the occurrence of quantum Mpemba effect and its role reversal in the dissipative Dicke model described by Eq.~\eqref{original_hamiltonian}, using differential quantum coherence, differential entanglement, and the trace distance between the time-evolved and steady states. Fig.~\ref{fig:Schematic} schematically illustrates our definitions of QME and its role reversal through an analogous real-life scenario.

\begin{figure*}
    \centering
    \includegraphics[
        width=18.5cm,height=7cm,
    ]{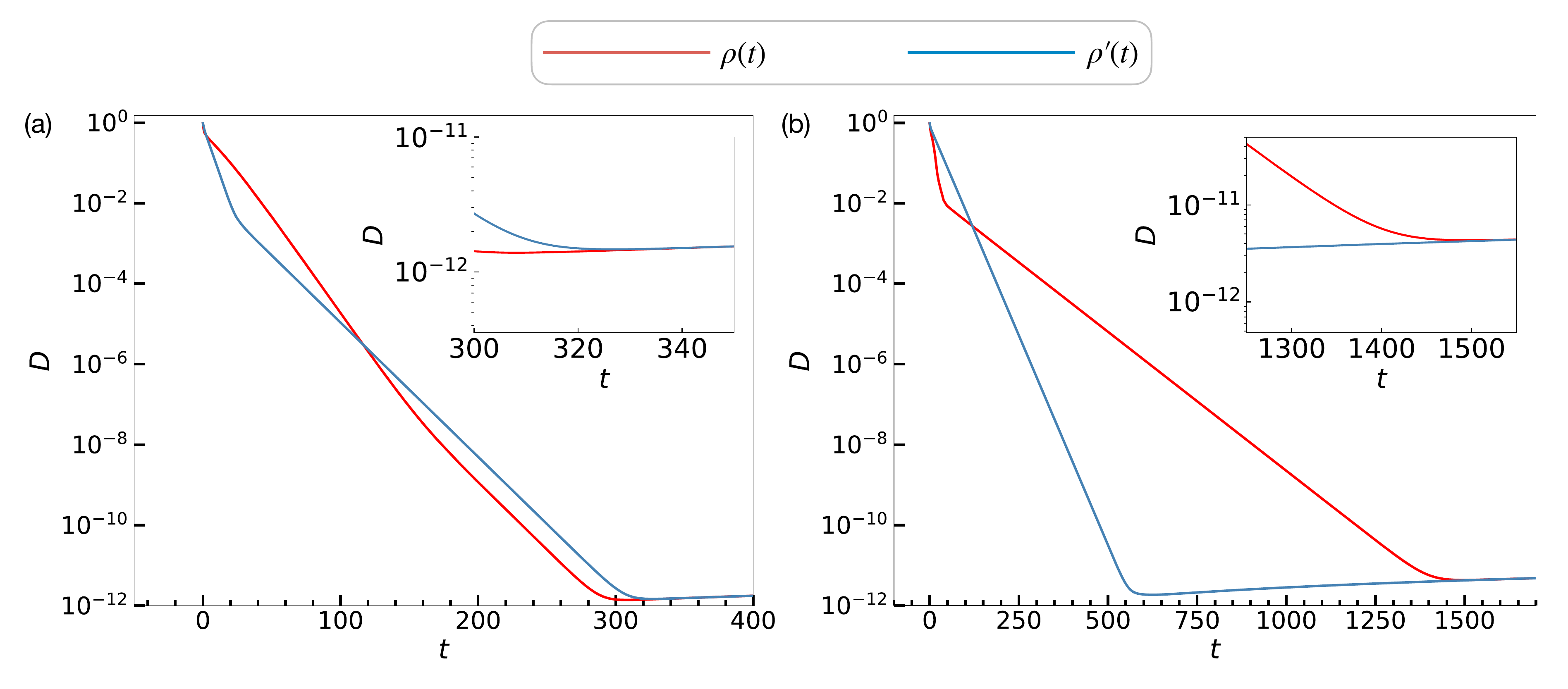}
    \caption{\textbf{Mpemba effect and role reversal characterized by trace-distance -based measure.} We plot the trace distance of time-evolved states $\{\rho(t),\rho'(t)\}$ at any time $t$ from the steady state $\rho_{ss}$. We consider $N=25$, $\Omega = 3, k = 1,$ and $ g=1$. In panel (a), when we choose $\omega = 1, $ $\rho'_0$ evolves faster than $\rho_0$ towards the steady state $\rho_{ss}$. Furthermore, we see that the two curves of the trace distance between time-evolved states and steady state corresponding to $\{\rho(t),\rho'(t)\}$, obtained in panel (a), interchange their roles for different value of $\omega$, i.e., $\omega = 0.1$ for the same pair of initial states $\{\rho_{0},\rho'_{0}\}$. Here the smooth red and blue lines correspond to the dynamics of trace distance of $\rho_0$ and $\rho'_0$ from the steady state in both the sub-figures (a) and (b). Therefore, panel (b) manifests the QME  and role reversal by panel (a), characterized by a trace-distance -based measure. In all the sub-figures, the time is described in the unit $\frac{\Omega}{\hbar}$.}
    \label{fig:Tr_dist_Mpemba_&_RR}
\end{figure*}

\subsection{Quantum coherence}
\label{subsec: coherence}

Here, we examine differential quantum coherence as a key indicator for identifying the
QME and demonstrate the emergence of role reversal in terms of
coherence. As the coherence of the steady state is zero for the dynamics considered here, the evolution of the differential coherence is identical to the evolution of the coherence. To this end, we focus on the qubit dissipative Dicke model, described
by Eq.~\eqref{original_hamiltonian} with $N=1$. We consider a pair of initial states
$\{\rho_0, \rho'_0\}$, where $\rho'_0 = U(\beta)\rho_0 U^{\dagger}(\beta)$, and the
unitary operator $U(\beta)$ is defined in Sec.~\ref{sIII}.

We express the qubit initial state $\rho_0$ in the Bloch-sphere representation, given in Eq.~\eqref{state_Bloch_sphere}. For the numerical analysis, we choose
$r_x = r_y = 0.4$ and $r_z = \sqrt{1 - (r_x^2 + r_y^2)}$. The Bloch component $r_z$ remains invariant under the action of the coherence-preserving
unitary operator, defined on the eigenbasis of $\sigma_z$. 
Throughout our study in this subsection, we fix the parameter of the coherence-preserving unitary to
$\beta = 0.33\pi$, which generates the transformed initial state $\rho'_0$.

To capture the occurrence of the QME, 
we consider the parameters of the dissipative Dicke model,  $\Omega = 1$, $k = 1$, $g = 1$, and $\omega = 0.1$. We then focus on the
coherence dynamics of the two systems prepared in the initial states $\rho_0$ and $\rho'_0$,
quantified using the $l_1$-norm of coherence. The states $\rho(t)$ and $\rho'(t)$ correspond to the time-evolved state at any time $t$ corresponding to the initial states $\rho_0$ and $\rho'_0$ respectively.
In Fig.~\ref{fig:RR_plots_resources} (a) and (b), the time evolution of the coherence for the
initial states $\rho_0$ and $\rho'_0$ is represented by the red and blue smooth curves,
respectively. It is evident that both initial states have the same initial coherence.
However, the coherence of $\rho(t)$ decays faster, reaching approximately zero (up to a
numerical precision of $10^{-4}$) around time $t \approx 8.79$.
On the other hand, the coherence of $\rho'(t)$ reaches the zero-coherence value (up to a
numerical precision of $10^{-4}$) at a later time, approximately $t \approx 10.08$, in
comparison to $\rho(t)$. Here, all the times is described in the unit of $\frac{\Omega}{\hbar}$. The faster decoherence of $\rho(t)$ than that of $\rho'(t)$ is more clearly depicted in the inset of Fig.~\ref{fig:RR_plots_resources} (b).
This behavior clearly shows the existence of QME
in terms of coherence, as one system relaxes to zero coherence faster than the other,
despite both having identical initial coherence.

To further examine the existence of role reversal, we keep the parameters $\Omega$ and
$k$ fixed at the same values as in Fig.~\ref{fig:RR_plots_resources}(a), while changing
$g$ from $1$ to $4.5$ and $\omega$ from $0.1$ to $1.0$. The numerical results for this
set of parameters are presented in Fig.~\ref{fig:RR_plots_resources} (b). In contrast to
Fig.~\ref{fig:RR_plots_resources} (a), a clear role reversal is observed: the coherence of
the system initially prepared in $\rho'_0$ now decays faster and reaches zero
asymptotically earlier than that of the system initially in $\rho_0$. This behavior is
more clearly highlighted in the inset of Fig.~\ref{fig:RR_plots_resources} (b), thereby
providing clear evidence of the role-reversal in QME phenomenon characterized by differential quantum coherence.


\subsection{Entanglement}
\label{subsec-entanglement}
In this subsection, we investigate the QME and the emergence of its role
reversal using another quantum resource, i.e., entanglement. We consider a composite system partitioned into two subsystems, labeled $A$ and $B$,
associated with the Hilbert spaces $\mathcal{H}_A$ and $\mathcal{H}_B$, each of
dimension $d_1$. 
The Hamiltonians governing subsystems $A$ and $B$ are described by Dicke models, denoted as $H^A$ and $H^B$, respectively. 
The total Hamiltonian of the
composite system is given by
\begin{align}
    H^{AB} = H^A \otimes I_{d_1} + I_{d_1} \otimes H^B, \nonumber
\end{align}
where the forms of $H^A$ and $H^B$ are written in Eq.~\eqref{original_hamiltonian} and $I_{d_1}$ denote the identity operators on both the
Hilbert spaces $\mathcal{H}_A$ and $\mathcal{H}_B$. Since the two
subsystems are non-interacting, the Lindblad jump operators for the composite system are given by ${L}_1^A \otimes I_{d_1}$ and $I_{d_1} \otimes L_1^B$.

We consider a pair of initial bipartite entangled states, $\rho_0$ and $\rho'_0$, related by a local unitary
transformation, such that both states possess the same initial
entanglement. Since the steady value of entanglement vanishes for the dynamics considered here, we focus directly on the time evolution of the entanglement of the pair of states rather than on the differential entanglement.

For the numerical analysis, we consider an equal number of spins in each subsystem,
namely $N_A = N_B = 3$, where $N_A$ and $N_B$ denote the number of spins in subsystems
$A$ and $B$, respectively.
 The Hamiltonian parameters for subsystems $A$ and $B$ are chosen as follows:
$\Omega_A = 3$, $g_A = 1$, $k_A = 1$, and $\omega_A = 1$ for subsystem $A$, and
$\Omega_B = 2.5$, $g_B = 3.5$, $k_B = 3$, and $\omega_B = 8.88$ for subsystem $B$.

Here, the subscripts $A$ and $B$ label the corresponding subsystems. We choose
$\rho_0 = |\psi\rangle\langle\psi|$, where
$|\psi\rangle = \frac{1}{\sqrt{3}} \sum_{i=0}^{2} |i_A i_B\rangle$
is a pure entangled state with respect to the bipartition $A\!:\!B$. The initial states $\rho_0$ and $\rho'_0$ are related by a local unitary transformation
$U = U_A \otimes U_B$, where the unitaries $U_A$ and $U_B$ are sampled randomly.
 We next investigate the entanglement dynamics using the logarithmic negativity
$\mathcal{L}_N$ as the entanglement measure. The states $\rho(t)$ and $\rho'(t)$ denote the time-evolved states at time $t$
originating from the initial states $\rho_0$ and $\rho'_0$, respectively.
In Fig.~\ref{fig:RR_plots_resources} (c) and (d),
the time evolution of entanglement for the initial states $\rho_0$ and $\rho'_0$ is
shown by the red and blue smooth curves, respectively, both starting from the same
initial entanglement. It is evident that the entanglement corresponding to $\rho(t)$
decays significantly faster than that of $\rho'(t)$, reaching zero entanglement (up to a
numerical precision of $10^{-18}$) at $t \approx 4.25$, after which it saturates. In
contrast, the entanglement associated with $\rho'_0$ vanishes at a later time,
approximately $t \approx 4.83$. The inset of Fig.~\ref{fig:RR_plots_resources} (c) provides a clearer view, highlighting the earlier saturation of the
entanglement corresponding to $\rho(t)$. Although both states possess identical initial
entanglement, their relaxation times differ, thereby manifesting the QME.

Furthermore, we numerically explore the occurrence of the role reversal in QME using differential entanglement. To
this end, we vary $\omega_B$ while keeping all other parameters and the set of initial states $\{\rho_0, \rho'_0\}$ fixed. Fig.~\ref{fig:RR_plots_resources} (d) shows that, for $\omega_B = 2.79$, the two curves interchange their roles: the entanglement of $\rho'(t)$ now decays faster than that of $\rho(t)$, in contrast to the behavior observed
in Fig.~\ref{fig:RR_plots_resources} (c), indicating a clear role reversal in the
entanglement dynamics.
The entanglement of $\rho(t)$ and $\rho'(t)$ vanishes to zero, within a numerical precision of order $10^{-18}$, at $t \approx 1.48$ and $t \approx 1.47$, respectively,
as clearly illustrated in the inset of Fig.~\ref{fig:RR_plots_resources} (d). In the numerical study all the times is described in the unit of $\frac{\Omega}{\hbar}$.
 Thus, the existence of the role reversal phenomenon in the context of QME is observed here, characterized by differential entanglement.


\subsection{Distance to steady state}
\label{subsec-tr_dist}
Here, we study the QME and its role reversal in terms of trace distance with the steady state $\rho_{ss}$ from the time-evolved states $\rho(t)$ and $\rho'(t)$, corresponding to the initial states $\rho_0$ and $\rho'_0$ respectively.
 We choose an arbitrary value of $\omega$. we take a randomly chosen initial pure state $\rho_{0}$ and then for that particular value of $\omega$, we construct the unitary operator $U$ according to the prescription, given in~\cite{Mpemba_Dicke_1_&_HS_dist_1} and obtain $\rho'_{0}=U\rho_{0}U^{\dagger}$, so that $\rho'(t)$ evolves faster than $\rho(t)$. Although the study of the QME and its role reversal in terms of quantum
resources was limited to small-sized systems, we have extended the analysis to larger system sizes using distance-based measures by considering $N=25$. For the numerical study, we consider $\Omega = 3, k = 1, g=1$, $\omega = 1 $, and constructed the unitary operator according to the prescription provided in~\cite{Mpemba_Dicke_1_&_HS_dist_1}, for the randomly chosen pure state $\rho_{0} = | \psi \rangle \langle \psi |$, so that $\rho'(t)$ evolves faster than $\rho(t)$ towards the steady state $\rho_{ss}$. Here the distance from any state at any time $t$ from the steady state is described with the help of the trace distance measure ($D$). The trace distance $D$ between two states $\rho_{1}$ and $\rho_{2}$ is defined as, $D \coloneqq \frac{1}{2} \Tr(\sqrt{A^{\dagger} A})$, where $A \coloneqq \rho_{1} - \rho_{2}$.

The trace distance between two states is invariant under unitary evolution, and the steady state $\rho_{ss}$ is unique for this setup, so the initial distance of $\rho_{ss}$ from $\rho_0$ and $\rho'_0$ is the same. It is well evident in Fig.~\ref{fig:Tr_dist_Mpemba_&_RR} (a) and (b), where the smooth red and blue curves represent the dynamics of trace distance of the time-evolved states $\rho(t)$ and $\rho'(t)$,  from the steady state respectively. Fig.~\ref{fig:Tr_dist_Mpemba_&_RR} (b), we see that $\rho'(t)$ reaches faster towards the steady state $\rho_{ss}$ up to the numerical precision of $10^{-12}$, spending less amount of time, i.e., $t \approx 524.5$. On the other hand $\rho_0$ evolves slower towards $\rho_{ss}$ up to that precision, spending much time, i.e., $t \approx 1347.5$. The inset of Fig.~\ref{fig:Tr_dist_Mpemba_&_RR} (b) provides a spectacular view that $\rho'_0$ evolves faster than $\rho_0$ towards $\rho_{ss}$, showcasing QME.

Now we vary $\omega$, keeping all other model parameters and also the pair of initial states $\{\rho_{0},\rho'_{0}\}$ fixed. For $\omega = 0.1 $, we see that the two curves interchange their roles for the same pair of initial states $\{\rho_{0},\rho'_{0}\}$. Fig.~\ref{fig:Tr_dist_Mpemba_&_RR} (a) exhibits that $\rho(t)$ reaches faster towards the steady state $\rho_{ss}$ up to the numerical precision of $10^{-12}$, spending less amount of time, i.e., $t \approx 262$. On the other hand, $\rho'_0$ evolves slower towards $\rho_{ss}$ up to that precision spending much time, i.e., $t \approx 281.5$. In this study also the unit of time is considered to be  $\frac{\Omega}{\hbar}$ as the previous cases. The inset of Fig.~\ref{fig:Tr_dist_Mpemba_&_RR} (a) clearly depicts that $\rho'_0$ evolves slower than $\rho_0$ towards $\rho_{ss}$. Thus, the two panels of Fig.~\ref{fig:Tr_dist_Mpemba_&_RR} showcase two scenarios of QME using trace distance between time-evolved state and steady state, with a reversed role of each other, for sufficiently large values of $N$, i.e., $N=25$.
 



\section{Conclusion}
\label{sec-Conclusion}
In conclusion, we analyzed the quantum Mpemba effect in the dissipative Dicke model, comprising a spin-1/2 ensemble coupled to a bosonic harmonic oscillator which in turn connected to a bosonic bath of harmonic oscillators. We analytically derived a sufficient condition for the occurrence of the quantum Mpemba effect in this model, characterized by differential quantum coherence. Next, we introduced and analyzed a phenomenon, role reversal in the quantum Mpemba effect, wherein changes in the Hamiltonian parameters invert the relaxation ordering of a given pair of initial states, causing the state that initially relaxes faster to become slower, and vice versa. 
Furthermore, we demonstrated the role reversal in the quantum Mpemba effect for this Dicke model using different classes of quantifiers, including physical-quantity–based measures such as differential quantum coherence and entanglement, as well as the distance-based measure, the trace distance between the time-evolved and steady states.

\section{ACKNOWLEDGMENTS}
 The research of
PC was supported by the INFOSYS scholarship.

\onecolumngrid
\section*{Appendix}

\label{sec-Appendix}
\appendix

\section{Detailed derivation of $l_1$-norm coherence under Lindblad Markovian dynamics} 
\label{Ap: l1_norm}
Here our aim is to evaluate the $l_1$ norm measure of $\rho(t)$ with the help of Eq.~\eqref{l1_norm}. The $(j,k)$-th element (where $j \neq k$) of $\rho(t)$, given by Eq.~\eqref{Lind_Markov_Evo} is
\begin{align*}
    \rho_{jk} &= \sum_{i=2}^{d^{2}}{e^{\lambda_{i}t}\frac{\Tr(l_{i}^{\dagger}\rho_{0})}{\Tr(l_{i}^{\dagger}r_{i})} (r_{i})_{jk}} ~[\because (\rho_{ss})_{jk} = 0 ~\text{for}~ j \neq k].
\end{align*}
Taking modulus in both sides, and then summing over $j$ and $k$ with $j \neq k$, in both sides, we obtain
\begin{align*}
    &\sum_{j \neq k}|\rho_{jk}| = \sum_{j \neq k}|\sum_{i=2}^{d^{2}}{e^{\lambda_{i}t}\frac{\Tr(l_{i}^{\dagger}\rho_{0})}{\Tr(l_{i}^{\dagger}r_{i})} (r_{i})_{jk}}|.\\
\end{align*}
As the LHS equals to $l_{1}(\rho)$ according to Eq.~\eqref{l1_norm}, hence Eq.~\eqref{coh} is derived.
Note: Eq.~\eqref{coh} holds only for the situations where the steady state $\rho_{ss}$ is diagonal on that basis.
\section{Relation between $c'_i$ and $c_i$}
\label{Ap: c_relation}
Let us consider the initial states $\rho_0$ and $\rho'_0$ written in the Bloch-sphere
representation as defined in Eq.~\eqref{state_Bloch_sphere}. The state $\rho'_0$ is obtained
from $\rho_0$ by applying a coherence-preserving unitary operation represented as
$U(\beta)=e^{i\beta\sigma_z}$, such that
$\rho'_0 = U(\beta)\rho_0 U^{\dagger}(\beta)$, and can be expressed as


\begin{align}
  \rho_0 = \frac{1}{2}[I_{2} + r_x \sigma_x+r_y \sigma_y +r_z \sigma_z],\hspace{0.12cm}   \rho'_0 = \frac{1}{2}[I_{2} + (r_x \cos (2 \beta )+r_y \sin (2 \beta )) \sigma_x + (r_y \cos (2 \beta )-r_x \sin (2 \beta )) \sigma_y + r_z \sigma_z] \label{rho0p}.
\end{align}

Here ${r_x,r_y,r_z}$ is the Bloch vector of $\rho_0$. Let subtracting $\rho'_0$ and $\rho_0$, and obtain

\begin{align*}
\rho'_0 - \rho_0 = \sin {\beta }[ (r_y \cos {\beta }-r_x \sin {\beta }) \sigma_{x} - (r_x \cos {\beta }+r_y \sin {\beta }) \sigma_{y}].
\end{align*}
Multiplying $l_{i}^{\dagger}$ from left and then evaluating trace in both sides, we obtain
\begin{align*}
    c'_i = c_i + \sin{\beta} [(r_y \cos{\beta} - r_x \sin{\beta}) (\alpha_x)_i - (r_x \cos{\beta} + r_y \sin{\beta}) (\alpha_y)_i],
~\text{where}~ (\alpha_x)_i = \Tr(l_{i}^{\dagger} \sigma_x) and (\alpha_y)_i = \Tr(l_{i}^{\dagger} \sigma_y).
\end{align*}
Hence, Eq.~\eqref{cpi} is derived.
\section{Proof of Theorem.~\ref{Theorem: suff_cond}}
\label{Ap: proof}

Here we have provided the detail proof of the theorem. \ref{Theorem: suff_cond}. Here, we consider the Bloch vector element of $r_x, r_y$ such that $r_x=r_y$.
We choose the parameters of the Dicke model such that
$\Omega = \omega = k \coloneqq p$, where $p > 0$.
We further define $q \coloneqq \sqrt{g^{4} - 25p^{4}}$, which is real for $g \geq \sqrt{5}\,p$.
 Then the matrix representations of the Hamiltonian $\tilde{H}$ and the Lindblad jump operator $\tilde{L}_1$ (described in \eqref{H_eff} and \eqref{eff_jump_operator}), are as the following:
\begin{align*}
    \tilde{H} =
    \begin{pmatrix}
        -\frac{g^2}{5p} + \frac{p}{2} & 0 \\
        0 & -\frac{g^2}{5p} - \frac{p}{2}\\
    \end{pmatrix},
    \qquad
    \tilde{L}_1 =
    \begin{pmatrix}
        0 & -\frac{(2+i)g}{5\sqrt{p}} \\
        -\frac{(2+i)g}{5\sqrt{p}} & 0\\
    \end{pmatrix}.
\end{align*}

Then we obtain the matrix representation of  $\mathcal{L}_{v}$, given by Eq.~\eqref{vec_liouv}, is given by 
\begin{align*}
    \mathcal{L}_{v} =
    \begin{pmatrix}
        -\frac{g^2}{5 p} & 0 & 0 & \frac{g^2}{5 p} \\
        0 & -\frac{g^2}{5 p}+i p & \frac{g^2}{5 p} & 0\\
        0 & \frac{g^2}{5 p} & -\frac{g^2}{5 p}-i p & 0\\
        \frac{g^2}{5 p} & 0 & 0 & -\frac{g^2}{5 p}
    \end{pmatrix}.
\end{align*}

Then we obtain the eigenvalues and right and left eigenmatrices of $\mathcal{L}_{v}$. The eigenvalues are, $\lambda_{1} = 0$, $\lambda_{2} = -\frac{2 g^2}{5 p}$, $\lambda_{3} = \frac{-\sqrt{g^4-25 p^4}-g^2}{5 p}$ and $\lambda_{4} = \frac{\sqrt{g^4-25 p^4}-g^2}{5 p}$.
We also notice that each of them has non-positive real part, manifesting the dissipative nature of Lindblad Markovian dynamics.
We also evaluate all the right eigenmatrices $r_i$'s and the left eigenmatrices $\ell_i$'s. We obtain that $r_1 = I_2$ which tells us, with the help of Eq.~\eqref{rho_ss}, that the steady state $\rho_{ss}$ is a zero-coherence state on the eigenbasis of $\sigma_z$.
Now we evaluate all the $k_i$'s, $c_i$'s and with the help of \eqref{cpi_s} $c'_i$'s, which are the provided in the following chart:
\begin{table}[H]
\centering
\begin{tabular}{|c|c|c|c|}
\hline
$k_1=2$ & $k_2=2$ & $k_3=1+\frac{\left(\sqrt{g^4-25 p^4}-5 i p^2\right)^2}{g^4}$ & $k_4 = 1+\frac{\left(\sqrt{g^4-25 p^4}+5 i p^2\right)^2}{g^4}$ \\ \hline
$c_1 = 1$ & $c_2 = -r_z$ & $c_3 = \frac{\left(\frac{1}{2}-\frac{i}{2}\right) \left(-i \sqrt{g^4-25 p^4}+g^2-5 p^2\right) r_x }{g^2}$ & $c_4 = \frac{\left(\frac{1}{2}+\frac{i}{2}\right) \left(\sqrt{g^4-25 p^4}-i g^2+5 i p^2\right) r_x }{g^2}$ \\ \hline
$c'_1 = 1$ & $c'_2 = -r_z$ & $c'_3 = \frac{\left(\frac{1}{2}-\frac{i}{2}\right) e^{-2 i \beta } \left(-i \sqrt{g^4-25 p^4}+g^2 e^{4 i \beta }-5 p^2\right) r_x }{g^2}$ & $c'_4 = \frac{\left(\frac{1}{2}+\frac{i}{2}\right) e^{-2 i \beta } \left(\sqrt{g^4-25 p^4}+(-i) g^2 e^{4 i \beta }+5 i p^2\right) r_x }{g^2}$ \\ \hline
\end{tabular}
\end{table}

Now, we evaluate the states under Lindblad Markovian evolution with the help of Eq.~\eqref{Lind_Markov_Evo}. The time evolved state $\rho(t)$ corresponding to the initial state $\rho_0$ is governed by
\begin{align*}
    \rho(t) =& \frac{1}{2}
    \begin{pmatrix}
        \left(r_z e^{-\frac{2 g^2 t}{5 p}}+1\right) & \frac{\left(1+i\right) r_x e^{-\frac{g^2 t}{5 p}} \left(q \sinh \left(\frac{t q}{5 p}\right)-i \left(g^2+5 p^2\right) \cosh \left(\frac{t q}{5 p}\right)\right)}{g^2+5 p^2}\\
        \frac{\left(1+i\right) r_x e^{-\frac{g^2 t}{5 p}} \left(\left(g^2+5 p^2\right) \cosh \left(\frac{t q}{5 p}\right)-i q \sinh \left(\frac{t q}{5 p}\right)\right)}{g^2+5 p^2} & 1- r_z e^{-\frac{2 g^2 t}{5 p}}
    \end{pmatrix}.
\end{align*}
Therefore the coherence of $\rho(t)$ in terms of $l_1$ norm is given by 
\begin{equation}
    l_{1}(\rho(t)) = |r_x| \sqrt{2} e^{-\frac{g^2 t}{5 p}} \left(\frac{\left(g^2 \cosh \left(\frac{2 t \sqrt{g^4-25 p^4}}{5 p}\right)+5 p^2\right)}{g^2+5 p^2}\right)^{1/2} \label{l1}.
\end{equation}
Similar to $\rho(t)$, the time-evolved state $\rho'(t)$ corresponding to the initial state $\rho'_0$ is governed by 

\begin{align*}
    \rho' (t) =
    \resizebox{\columnwidth}{!}{$
    \begin{pmatrix}
        \frac{1}{2} \left(r_z e^{-\frac{2 g^2 t}{5 p}}+1\right) & \frac{\left(\frac{1}{2}+\frac{i}{2}\right) e^{-2 i \beta } r_x \left(\frac{\left(q+(-i) g^2 e^{4 i \beta }+5 i p^2\right) e^{\frac{t \left(q-g^2\right)}{5 p}}}{1+\frac{\left(q+5 i p^2\right)^2}{g^4}}-\frac{i \left(-i q+g^2 e^{4 i \beta }-5 p^2\right) e^{-\frac{t \left(q+g^2\right)}{5 p}}}{1+\frac{\left(q-5 i p^2\right)^2}{g^4}}\right)}{g^2}\\
        \frac{\left(\frac{1}{4}+\frac{i}{4}\right) r_x e^{-\frac{t \left(q+g^2\right)}{5 p}-2 i \beta } \left(q+\left(q+5 i p^2\right) e^{\frac{2 t q}{5 p}}+i g^2 e^{4 i \beta }+(-i) g^2 e^{\frac{2 t q}{5 p}+4 i \beta }-5 i p^2\right)}{q} & \frac{1}{2}-\frac{1}{2} r_z e^{-\frac{2 g^2 t}{5 p}}
    \end{pmatrix}.
    $}
\end{align*}
The coherence  of $\rho'(t)$ is given by
\begin{align}
    &l_{1}(\rho'(t)) = \sqrt{2} |r_x|  e^{-\frac{g^2 t}{5 p}} \frac{\sqrt{g^4 \cosh \left(\frac{2 t q}{5 p}\right)+g^2 \left(q \sin (4 \beta ) \sinh \left(\frac{2 t q}{5 p}\right)-10 p^2 \cos (4 \beta ) \sinh ^2\left(\frac{t q}{5 p}\right)\right)-25 p^4}}{q}. \label{lp1}
\end{align}
Now we evaluate the difference of $l_1$-norm between the two time-evolved states $\rho'(t)$ and $\rho(t)$ is obtained by subtracting \eqref{lp1} from \eqref{l1} and and is expressed as
\begin{align}
l_{1}(\rho'(t))-l_{1}(\rho(t))
=\sqrt{2} |r_x| e^{-\frac{g^2 t}{5 p}} \left(\sqrt{a}-\sqrt{b}\right). \label{l1_diff}
\end{align}
where $a \coloneqq \frac{g^4 \cosh (\frac{2 t q}{5 p})+g^2 \left(q \sin (4 \beta ) \sinh (\frac{2 t q}{5 p}\right)-10 p^2 \cos (4 \beta ) \sinh ^2\left(\frac{t q}{5 p}\right))-25 p^4}{g^4-25 p^4}$ and $b \coloneqq \frac{ (g^2 \cosh \left(\frac{2 t q}{5 p}\right)+5 p^2)}{g^2+5 p^2}$.

Here, we focus on identifying the region in the parameter space of the coherence-preserving unitary parameter $\beta$ and the Hamiltonian parameters for which the Mpemba effect manifests. Specifically, we consider two systems initialized in the states $\{\rho_0,\rho'_0\}$ that possess the same differential coherence (which is the difference of coherence between the initial state and steady state), while the differential coherence of $\rho_0$ relaxes rapidly to its zero-coherence value. 
Note that, since the steady state $\rho_{\mathrm{ss}}$ is an incoherent state, the differential coherence at any time during the evolution reduces to the coherence of the state itself. Therefore, it suffices to track the time evolution of the coherence of the states.
 Our objective is to establish a condition under which, for all finite times $t \in (0,\infty)$, the inequality
$l_{1}\bigl(\rho'(t)\bigr) - l_{1}\bigl(\rho(t)\bigr) > 0
$
holds.
From Eq.~\eqref{l1_diff}, noting that
$
|r_x|\sqrt{2}\, e^{-\frac{g^2 t}{5p}} > 0$, $t \in (0,\infty),
$
it follows that
$
l_{1}\bigl(\rho'(t)\bigr) - l_{1}\bigl(\rho(t)\bigr) > 0
$
for all finite times if and only if
$
\sqrt{a} - \sqrt{b} > 0.
$
The quantity $\sqrt{a}-\sqrt{b}>0$ when $a>b$.
The difference between $a$ and $b$ in terms of $\beta$, $q$, $p$ and $t$ is given by,
\begin{align}
a-b = \frac{g^2 \left(q \sin (4 \beta ) \sinh \left(\frac{2 t q}{5 p}\right)+20 p^2 \sin ^2(2 \beta ) \sinh ^2\left(\frac{t q}{5 p}\right)\right)}{q^2} \label{core_positivity}.
\end{align}
Now, since $g^2>0$, $q^2>0$, $20p^2\sin^2(2\beta)\sinh^2\!\left(\frac{tq}{5p}\right)>0$, and
$
q\,\sinh\!\left(\frac{2tq}{5p}\right)>0 $ for $t>0$,
it follows that Eq.~\eqref{core_positivity} is strictly positive for all finite times
when $\beta \in \left(0,\frac{\pi}{4}\right]\cup\left(\frac{\pi}{2},\frac{3\pi}{4}\right] \cup (\pi,\frac{5\pi}{4}] \cup (\frac{3\pi}{4},\frac{7\pi}{4}]$, where $\sin(4\beta)>0$, which renders the expression in Eq.~\eqref{core_positivity} positive. 


Hence, we see that $l_1(\rho'(t))>l_1(\rho(t))$ for all finite time $t \in (0,\infty)$ when $\beta \in (0,\frac{\pi}{4}] \cup (\frac{\pi}{2},\frac{3\pi}{4}] \cup (\pi,\frac{5\pi}{4}] \cup (\frac{3\pi}{4},\frac{7\pi}{4}]$ for the initial states with $r_x=r_y$. Therefore, $\beta \in (0,\frac{\pi}{4}] \cup (\frac{\pi}{2},\frac{3\pi}{4}] \cup (\pi,\frac{5\pi}{4}] \cup (\frac{3\pi}{4},\frac{7\pi}{4}]$ with $r_x=r_y$ serves as a sufficient condition which assures that the time evolved state $\rho'(t)$ corresponding to the initial state $\rho'_0$ decoheres slower towards the zero coherence state than the time evolved state $\rho(t)$ corresponding to the initial state $\rho_0$ exhibiting quantum Mpemba effect.

Hence, the proof is completed.


\twocolumngrid
\bibliography{citations.bib}
\end{document}